\documentclass[aps,pra,twocolumn,showpacs]{revtex4}
\usepackage{mathrsfs}
\usepackage{amsmath}
\usepackage{graphics, graphicx,epsfig}
\usepackage{dcolumn}
\usepackage{amssymb}
\usepackage{bm}
\begin{document}
\title{ Detecting unambiguously non-Abelian geometric phases with trapped ions}
\author{Xin-Ding Zhang$^{1}$}
\author{Z. D. Wang$^{2}$}
\author{Liang-Bin Hu$^1$}
\author{Zhi-Ming Zhang$^3$}
\author{Shi-Liang Zhu$^{1}$}
\email{slzhu@scnu.edu.cn} \affiliation{$^1$Institute for Condensed
Matter Physics, School of Physics and Telecommunication Engineering, South China Normal University, Guangzhou, China \\
$^{2}$Department of Physics \& Center of Theoretical and
Computational Physics, University of Hong Kong, Pokfulam Road,
Hong Kong, China\\ $^3$Laboratory of Photonic Information
Technology, School of Information and Photoelectronics, South
China Normal University, Guangzhou, China}

\begin{abstract}
We propose for the first time an experimentally feasible scheme to
disclose the noncommutative effects induced by a light-induced
non-Abelian gauge structure with trapped ions. Under an appropriate
configuration, a true non-Abelian gauge potential naturally arises
in connection with the geometric phase associated with two
degenerated dark states in a four-state atomic system interacting
with three pulsed laser fields. We show that the population in
atomic state at the end of a composed path formed by two closed
loops $C_1$ and $C_2$ in the parameter space can be significantly
different from the composed counter-ordered path. This population
difference is directly induced by the noncommutative feature of
non-Abelian geometric phases and can be detected unambiguously with
current technology.
\end{abstract}

\pacs{03.65.Vf, 42.50.Gy}
\maketitle

\newpage


Gauge structures have been used to describe almost all the
fundamental interactions in nature and have been shown to play a
key role in modern physics. Gauge fields  are usually classified
as Abelian and non-Abelian ones according to the commutative rule
of the associated group elements. Remarkably, they also arise
naturally in the adiabatic evolution of simple quantum systems,
whose initial formulation has no apparent relationship to gauge
fields\cite{Berry,Wilczek}. Berry demonstrated that the adiabatic
cyclic evolution of a non-degenerated state leads to a scalar
geometric phase factors corresponding to a $U(1)$ Abelian gauge
field \cite{Berry}. Wilczek and Zee further generalized this idea
to the case of degenerate states and showed that a non-Abelian
structure emerges if a set of quantum states remains degenerate as
the Hamiltonian varies. When a system with $N$-fold degenerate
levels undergos cyclic process, the mapping corresponds to a
geometric phase factor with $SU(N)\times U(1)$ matrix
form\cite{Wilczek,Zee}. So far, gauge structures in such simple
quantum systems have found many applications in diverse fields in
physics\cite{Shapere,Kwiat,zhu,Satija, Ruseckas,
Osterloh,zanardi,Sjoqvist,duan, xdzhang, liyong, recati}.

Following the progress achieved in theory and application, direct
experimental observations of gauge structures in quantum systems
attracts increasing interests. Although Abelian geometric phases
have been tested experimentally by various means
\cite{Shapere,Kwiat}, the noncommutative effect, a key property of
non-Abelian structures has not been experimentally demonstrated.
In an interesting experiment\cite{tycko}, Tycko demonstrated the
effect of geometric phase on the magnetic resonance spectrum of
two-fold degenerated states\cite{tycko}. However, as pointed out
in Ref. \cite{Zee}, because of the experimental limitation in
nuclear magnetic resonance (NMR), only the (commutative) Abelian
part of the non-Abelian gauge structure has been experimentally
observed and the direct observation of noncommutativity of
non-Abelian gauge structures in NMR system is really difficult.
Recent experiments in trapped-ion systems provide high-precise
manipulations and measurements on  internal states of atoms
\cite{Wineland,Monroe1,Barton, Monroe, Ozeri, Kreuter, Rabl}.
Comparing with NMR systems, true non-Abelian structures can be
easier to be realized by modulating controlling parameters in
principle. Thus it is significant and important to explore whether a
trapped-ion system is promising for detecting the fundamental
non-Abelian characteristic.

In this paper, we analyze in detail why non-Abelian structures have
not been observed in NMR experiments, and propose an experimentally
feasible scheme to directly detect the observable effect induced by
the noncommutative property of the non-Abelian gauge potentials in a
trapped-ion system\cite{Note}. We show that a true non-Abelian gauge
potential naturally arises in connection with the geometric phase
associated with two degenerated dark states in a four-state atomic
system interacting with three pulsed laser fields.
The system under consideration is a four-level atom interacting with
resonant external laser pulses. Under an appropriate configuration,
there exist two-fold degenerated dark states and then a non-Abelian
gauge potential naturally arises. We design two specific closed
loops $C_1$ and $C_2$ in the parameter space, which generate two
different geometric phase factors $U_{1}$ and $U_{2}$, respectively.
Consider the composite path in which one traverses first $C_1$ and
then $C_2$. After a cycle, a phase factor of $U=U_2 U_1$ is
generated. In contrast, the other phase factor of $U'=U_1 U_2$ is
accumulated if one traverses first $C_2$ and then $C_1$. We find
that the population of the atoms in a specific state at the end of
the two evolutions $U$ and $U'$ can be significantly different. For
instance, the population difference can be as high as $43\%$ for the
typical experimental parameters. Note that the observe precision of
an atomic state can be higher than $99\%$, and thus the proposed
noncommutative effect can be detected unambiguously. Since the
variations of the parameters associated with the non-Abelian
geometric phases can also be easily manipulated via the laser beams,
the proposed scheme is very promising for detecting the fundamental
non-Abelian gauge structures.

We start by reviewing a general framework of non-Abelian adiabatic
geometric phase \cite{Wilczek,Zee} and its possible detection in
nuclear quadrupole resonance\cite{tycko}. Consider a family of
Hamiltonians $H(\chi_\mu)$ ($\mu=1,2,\cdots,n$) depending
continuously on parameters $\chi_\mu$, all of which have a set of
$N$ degenerate levels. Let $\eta_a$ be an $N$-fold degenerate set of
orthonormal instantaneous eigenstates of the Hamiltonian $H$
described by $H|\eta_a\rangle=E|\eta_a\rangle$ with $E$ being the
eigenvalue.
Under the adiabatic condition the instantaneous state
$|\psi_a\rangle$ will not overflow the state vector space spanned by
$\eta_{a}$. So $|\psi_a\rangle$ can always be expanded as a
superposition of $\vert \eta_a\rangle$: $\vert
\psi_{a}\rangle=\Sigma_{b}\vert \eta_{b}\rangle U_{ba}$.
Substituting the wave function $|\psi_a\rangle$ into the
time-dependent Schrodinger equation $i\frac{d}{dt}\vert
\psi_a\rangle=H\vert \psi_a\rangle$, we have
\begin{equation}
\dot{U}_{ba}=-\sum_{c} \langle \eta_{b}\vert\dot{\eta}_{c}\rangle
U_{ca},
\end{equation}
where a trivial overall dynamical phase term has been dropped. A
gauge structure appears when the Hamiltonian $H$ varies as the
parameters $\chi_{\mu}(t)$  vary slowly with time $t$. We can define
a gauge potential
\begin{equation}
\label{Potential} A_{ab\mu}=\langle \eta_{a}\vert
\frac{\partial}{\partial \chi^{\mu}}\vert \eta_{b}\rangle,
\end{equation}
and  have $\langle
\eta_{a}\vert\dot{\eta}_{b}\rangle=\Sigma_{\mu}A_{ab\mu}(d\chi^{\mu}/dt)$.
Integrating Eq. (2),  it is straightforward to obtain
\begin{equation}\label{U}
U_{ab}=\left[\mathcal {P}exp\left(-\int
A_{\mu}d\chi^{\mu}\right)\right]_{ab},
\end{equation}
where $\mathcal{P}$ denotes the path-ordered operator. The quantity
$A$ defined in Eq.(\ref{Potential}) indeed plays the role of a gauge
potential and the gauge group corresponds to the unitary freedom in
choosing the basis states $\eta_a$ \cite{Wilczek}. An intriguing
feature of the gauge potential $A$ lies in that it depends only on
the geometry of executed path in the space of degenerate levels.

In an earlier experiment, Tycko reported the effect of geometric
phase on the nuclear-quadrupole-resonance spectra\cite{tycko}. The
effective Hamiltonian describing Tycko's experiment is  given by
\begin{eqnarray}
H&=&(\mathbf{S}\cdot\mathbf{B})^{2}\notag\\
&=&(S_{x}\sin\theta\cos\varphi+ S_{y}\sin\theta\sin\varphi+
S_{z}\cos\theta)^{2}B^{2}, \label{H_exp}
\end{eqnarray}
where $\mathbf{S}$ denotes the nuclear spin operator for the
sample used in the experiment (a spin-$\frac{3}{2}$ $Cl$ atom in a
$NaClO_3$ crystal is considered), and $\mathbf{B}$ represents an
external magnetic field. There always exist a pairwise degeneracy
of states in the parameter space because the Hamiltonian is
invariant under the operation $\mathbf{S} \rightarrow
-\mathbf{S}$, thus non-Abelian gauge structure may appear in the
system. To address clearly the gauge structure, one can
parameterize the Hamiltonian (\ref{H_exp}) as follows
\begin{equation}\label{H}
H=B^{2}e^{-i\varphi S_{z}}e^{-i\theta S_{y}} S_{z}^{2} e^{i\theta S_{y}}e^{i\varphi S_{z}}.
\end{equation}
As a result, the instantaneous eigenstates can be directly written
as
\begin{equation}
\label{Basis} \vert\eta_{a}\rangle=e^{-i\varphi S_{z}}e^{-i\theta
S_{y}}|a\rangle,
\end{equation}
where $\vert a\rangle$ is one of the atomic state in the set $\{
|\pm \frac{3}{2}\rangle,|\pm\frac{1}{2}\rangle\}$. Here the state
$|m\rangle$ represents the eigenstates of the spin operator $S_z$
defined by $S_z|m\rangle=m|m\rangle$ and the states $|\pm m\rangle$
form a doubly degenerate sector. Substituting Eq.(\ref{Basis}) into
Eq.(\ref{Potential}), the corresponding gauge potentials
are given by
\begin{eqnarray}
A_{ab\varphi}&=&(-i) \langle a \vert(cos\theta S_{z}-sin\theta S_{x})\vert b \rangle,\notag\\
A_{ab\theta}&=&(-i) \langle a \vert S_{y}\vert b\rangle.
\end{eqnarray}
As clearly shown in Ref.\cite{Zee}, an Abelian structure occurs for
$|m|=\frac{3}{2}$ sector although those states are doubly
degenerate, while a true non-Abelian structure may appear for the
sector $|m|=\frac{1}{2}$. Explicitly, the gauge potentials for the
later case is given by,
\begin{equation}\label{A}
A=-i\left[\left(\frac{1}{2}\sigma_{z}\cos\theta -
\frac{\alpha}{2}\sigma_{x}\sin\theta\right) d\varphi +
\frac{\alpha}{2}\sigma_xd\theta\right],
\end{equation}
where $\alpha=S+1/2$ and $\sigma_{x,y,z}$ are Pauli matrixes.

To disclose unambiguously the non-Abelian characteristic of the
geometric phase, we must detect the physically observable effects
induced by the noncommutativity of the gauge structure. To this end,
here we consider two closed loops $C_{1}$ and $C_{2}$, both starting
and ending at the same point $\chi_0$ in the parameter space, where
the two corresponding evolution operators are denoted as $U_{1}$ and
$U_{2}$, respectively. For the composite path that the system first
traverses along loop $C_{1}$ and then $C_{2}$, the total phase
factor of $U=U_{2}U_{1}$ is generated. If the counter-ordered
evolution is executed, i.e., an evolution with first loop $C_{2}$
and then $C_{1}$, the phase factor generated is then given by that
of $U^{'}=U_{1}U_{2}$. In a general experiment, if the observable
effects are different for the phase factors of $U$ and $U^{'}$, then
Non-Abelian geometric structures are confirmed there. Otherwise,
only the Abelian part of the gauge structure is observed even if the
quantum states of the system are degenerate.

From the above discussion, it is easy to understand that only the
Abelian part of the gauge structures has been experimentally
observed in the Tycko's experiment, as has been addressed in
Ref.\cite{Zee}. In the Tycko's experiment, the angle $\theta$ is
fixed as a constant ($\cos\theta=1/\sqrt{3}$). It is due  to that
only the rotation around a fixed axis can be implemented fast enough
to generate observable phase shift in NMR systems. In this case
gauge structure $A$ is proportional to a fixed matrix $\cos \theta
\frac{\sigma_{3}}{2}-\alpha \sin\theta \frac{\sigma_{1}}{2}$ and the
evolution operator can be written as
$U=\exp[-i\varphi(\sigma_z/\sqrt{3}-\sigma_x)]$. Thus the $U$ and
gauge potential $A$ in different time sequences are always
commutable. From this point of view, we say that the gauge potential
loses its non-Abelian characteristic and is of Abelianization. In
order to demonstrate the non-Abelian characteristic in nuclear
quadrupole resonance, the system should traverse a non-Abelian path
in which both $\theta$ and $\varphi$ vary with time. Unfortunately,
it is difficult in an NMR setup.

\begin{figure}[tbhp]
\includegraphics[scale=0.8]{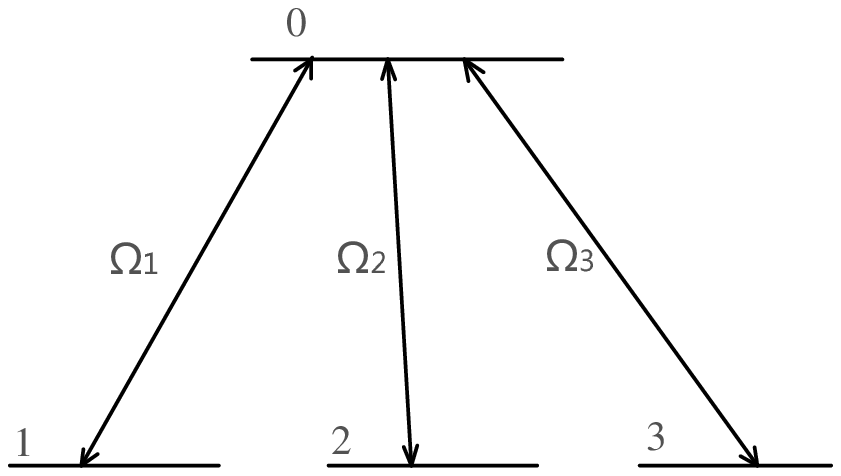}
\caption{Schematic map of the four-level system interacting with
resonant laser fields. States $\vert 0\rangle$ and $\vert
i\rangle$ $(i=1,2,3)$ are coupled by a resonant external laser
pulse with effective Rabi frequency $\Omega_{i}$.}
\end{figure}

Recently, with the rapid experimental progress in trapped ions
\cite{Wineland,Monroe1,Barton, Monroe, Ozeri, Kreuter, Rabl}, it
is quite possible to experimentally observe the non-Abelian
characteristic of gauge potentials with current technology. We now
start to present a scheme to detect the non-Abelian characteristic
of the gauge potentials in such system. The system under
consideration is an ion (or a set of ions) confined in a (linear)
Paul trap. Each ion has three ground (or metastable) states
$|1\rangle$, $|2\rangle$, $|3\rangle$, and one excited state
$|0\rangle$. The ground states could be different hyperfine levels
or in the same manifold but with different Zeeman sublevels, and
they are coupled to the excited state $|0\rangle$ by resonant
classical lasers with Rabi frequencies $\Omega_1$, $\Omega_2$ and
$\Omega_3$, respectively, as shown in Fig.1\cite{duan}. The
effective Hamiltonian for each ion in the rotating frame reads
\begin{equation}
\label{H_I}
  H_{I}=-\hbar(\Omega_{1}|0\rangle\langle 1|+\Omega_{2}|0\rangle\langle 2|+\Omega_{3}|0\rangle\langle 3|)+H.c.
\end{equation}
In a general case the Rabi frequencies $\Omega_\mu$ can be
parameterized with angle and phase variables according to
$\Omega_{1}=\Omega\sin\theta \cos\phi e^{i\varphi_{1}}$,
 $\Omega_{2}=\Omega\sin\theta \sin\phi e^{i\varphi_{2}}$, and $\Omega_{3}=\Omega \cos\theta e^{i\varphi_{3}}$, where
 $\Omega=\sqrt{|\Omega_1|^2+|\Omega_2|^2+|\Omega_3|^2}$.
Two adiabatic eigenstates of $H_{I}$ with zero eigenvalue (dark
state) are then found to be
$$
\vert D_{1}\rangle =\sin\phi e^{i\varphi}\vert 1\rangle -\cos\phi
e^{i\varphi}\vert 2\rangle,
$$
\begin{equation}
\label{Dark} \vert D_{2}\rangle =\cos\theta\cos\phi
e^{i\varphi}\vert 1\rangle + \cos\theta\sin\phi e^{i\varphi}\vert
2\rangle-\sin\theta|3\rangle,\end{equation} where we have assumed
that the laser phases satisfy the relations
$\varphi=\varphi_3-\varphi_1=\varphi_3-\varphi_2$. Substituting
 Eq.(10) into the
 formula (2), we have,
\begin{eqnarray}
A_{\theta}&=&0,         \notag\\
A_{\phi}&=&\left(
               \begin{array}{cc}
                 0 & -\cos\theta \\
                 \cos\theta & 0 \\
               \end{array}
             \right),          \notag\\
A_{\varphi}&=& \left(
                 \begin{array}{cc}
                   i & 0 \\
                   0 & i\cos^{2}\theta \\
                 \end{array}
               \right)
              .
\end{eqnarray}
Then it is straightforward to find that the gauge potentials are
given by
\begin{eqnarray}
A&=&\Sigma_{\mu}A_{\mu} d\chi^{\mu} \notag\\
 &=&i\left(\frac{1+\cos^2\theta}{2}I
 +\frac{sin^2\theta}{2}\sigma_z\right)d\varphi-i\sigma_y\cos\theta
 d\phi,\label{A_fomular}
 \end{eqnarray}
where $I$ is a $2\times 2$ unit matrix. Similar to the above
analysis in NMR system, the non-Abelian characteristic occurs only
when both the angle $\phi$ and phase $\varphi$ vary with time. For
instance,  if the phase  $\varphi$ is fixed, we have
\begin{equation}
U=\exp({i\gamma\sigma_y)}=\left( \begin{array}{cc}
                 \cos\gamma & \sin\gamma \\
                 -\sin\gamma & \cos\gamma \\
               \end{array}\right).\notag
\end{equation}
where $\gamma=\int \cos\theta d\phi$. This evolution and its
geometric phase  have been adressed in Ref. \cite{Unanyan1999}.
However, it is clear that the non-Abelian characteristic
(noncommutativity) of the potential $A$ is lost in this specific
case.  Therefore, rather than non-Abelian geometric phase factor,
the scheme proposed in \cite{Unanyan1999} can only be used to
observe the Abelian part of the geometric phase, although the
two-fold dark states remain degenerate as the Hamiltonian varies
with time.


To observe the true non-Abelian structure, we propose two specific
non-Abelian closed loops $C_1$ and $C_2$ in which both the phase
$\varphi$ and angle $\phi$ vary with time. In the closed loop $C_1$,
we assume that the Rabi frequencies are given by
\begin{eqnarray}
&&\Omega_{1}=\Omega_{0}f(t),\notag\\
&&\Omega_{2}=\Omega_{0}f^{2}(t),\notag\\
&&\Omega_{3}=\Omega_{0}e^{-t^2/\tau^{2}}e^{i\varphi},
\end{eqnarray}
where $\varphi=\frac{\pi}{\tau} t$, and $f(t)$ is set as
\begin{eqnarray}
f(t)=\left\{
              \begin{array}{c}
                \cos(\frac{\pi t}{2\tau}), \text{\ \ \ \ \ \ \ \ \ } -\tau\leq t \leq \tau, \\
                0  \text{, \ \ \ \ \ \ \ \ \ \ \ \ \ \       otherwise.}\\
              \end{array} \right.
\end{eqnarray}
In this case, the parameters $\theta(t)$ and $\phi(t)$ are
determined by
\begin{eqnarray} \label{A1}
&&\tan\phi(t)=\frac{|\Omega_{2}|}{|\Omega_{1}|}=\cos\left(\frac{\pi t}{2\tau}\right),\notag\\
&&\tan
\theta(t)=\sqrt{\frac{|\Omega_{1}|^2+|\Omega_{2}|^2}{|\Omega_{3}|^2}}=
\sqrt{\frac{\cos^{2}\left(\frac{\pi
t}{2\tau}\right)+\cos^{4}\left(\frac{\pi t}{2\tau}\right)}
{  \exp(  -\frac{2t^2} {\tau^2}  )}}.\notag\\
\end{eqnarray}
From time $-\tau \rightarrow \tau$, the parameters ($\theta, \phi,
\varphi$) vary from $(0, 0, -\pi)$ to $(0, 0, \pi)$, and it just
accomplishes a closed path in the parameter space. It is remarkable
that all parameters ($\theta, \phi, \varphi$) are completely
independent of the laser parameters $\Omega_0$ and $\tau$,
 provided that the adiabatic approximation is reasonable, thus it
is easy to design the required laser pulses. By substituting Eq.
(\ref{A1}) into Eq. (\ref{A_fomular}), the gauge potential $A_{1}$
generated by the loop $C_1$ can be derived, and  the associated
evolution operator is given by
 \begin{equation} \label{U1} U_{1}=\mathcal {P}
exp\left(-\oint_{c_{1}}A^{\mu}_{1}d\chi^{\mu}\right).
\end{equation}
We set the Rabi frequencies for the loop $C_2$ as follows,
\begin{eqnarray}
&&\Omega_{1}=\Omega_{0}f(t),\notag\\
&&\Omega_{2}=\alpha\Omega_{0}f^{2}(t),\notag\\
&&\Omega_{3}=\Omega_{0}e^{-(t-\beta\tau)^2/\tau^{2}}e^{i\varphi},
\end{eqnarray}
where $\alpha$ and $\beta$ are two variables to differentiate the
loops $C_{1}$ and $C_{2}$, and $\beta$ can also be considered as a
time delay among pulses $\Omega_{3}$ and $\Omega_{1}$, $\Omega_{2}$.
In this loop both the start and end points in the parameter space
($\theta, \phi, \varphi$) are the same as those in the loop $C_1$.
Similarly, a gauge potential $A_{2}$ will be generated in this loop
$C_2$, and the associated evolution operator is given by
\begin{equation}\label{U2}
U_{2}=\mathcal {P} exp\left(-\oint_{c_{2}}A^{\mu}_{2}d\chi^{\mu}\right).
\end{equation}
The analytical result of integral in Eq. (\ref{U1}) (or Eq.(
\ref{U2})) is difficult in a general case, because each segment in
the time-ordered exponential loop integral does not commute with the
next. Nevertheless, the integral can be evaluated easily by
numerical methods.

We now start to show that the composite path formed by the loops
$C_1$ and $C_2$ has noncommutative feature. It is straightforward to
check that the initial eigenstates $|D_{1,2}\rangle_i$ and the final
eigenstates $|D_{1,2}\rangle_f$ for both pathes $C_1$ and $C_2$ are
given by $|D_{1}\rangle_{i}=|D_{1}\rangle_{f}=-|2\rangle$ and
$|D_{2}\rangle_{i}=|D_{2}\rangle_{f}=|1\rangle$. We may initially
prepare the system as $|\Psi\rangle_{i}=|D_{2}\rangle_i$. Then we
let the effective Hamiltonian undergo a  composite path, first
$C_{1}$ and then $C_{2}$. The total evolution operator is a
two-by-two matrix denoted as $U=U_{2}U_{1}$. After this evolution,
the final state becomes
\begin{equation} | \Psi\rangle_{f}=U_{12}| D_{1}\rangle_{f}+U_{22}|
D_{2}\rangle_f= -U_{12}\vert 2\rangle+U_{22}\vert
1\rangle,\end{equation}
where $U_{12}$, $U_{22}$ are the elements
of the matrix U. On the other hand, if we let the Hamiltonian
tracks the counter-ordered path, i.e., first $C_{2}$ and then
$C_{1}$, the final state becomes
\begin{equation}|\Psi^{'}\rangle_{f}= -U^{'}_{12}\vert
2\rangle+U^{'}_{22}\vert 1\rangle,\end{equation} where $U_{12}^{'}$
and $U_{22}^{'}$ are the elements of  $U^{'}=U_{1}U_{2}.$ Denote
$P=|U_{22}|^{2}$ and $P^{'}=|U^{'}_{22}|^{2}$, which correspond to
the possibilities to find the final state in the state $\vert
1\rangle$ after composite paths $C_{2}C_{1}$ and $C_{1}C_{2}$,
respectively, the population difference $P_d$ between the two
composite paths is given by
\begin{equation}
\label{P_d} P_d=P'-P=|U^{'}_{22}|^{2}-|U_{22}|^{2}.
\end{equation}
The most essential feature of non-Abelian gauge structure (the
noncommutativity character) can be unambiguously
 observed if the population difference $P_d\not= 0.$

 To  show clearly the observable effect induced by the noncommutativity
 feature of the non-Abelian geometric phases,
we here numerically calculate the population difference described by
Eq.(\ref{P_d}) for some typical parameters. The populations $P$,
$P'$, and the difference $P_d$ between them as functions of the
parameters $\alpha$ and $\beta$ are plotted in Fig. 2 (a) and (b).
In the calculations for the evolution $U$ $(U')$, we let $U_1$
($U_2$) implement during time $-\tau \to \tau$, while $U_2$ $(U_1)$
implement from $\tau$ to $3\tau$. The effects induced by the
noncommutative property are clear since the difference $P_d$ can be
non-zero. To have a quantitative idea about the difference between
the evolution $U$ and $U'$, the population difference $P_d$ as a
function of $\beta$ for $\alpha=7$ is shown in Fig.2(c). In this
case $P_d= 43.2\%$ is derived for $\beta=0.5$.

  Nowadays the measurement technique of
population of atomic states has been well developed and the
detection precision can be higher than $99\%$. This high-precision
measurement  really makes it  experimentally feasible to directly
``see'' the noncommutativity  of non-Abelian gauge potentials by
measuring the populations of atomic states.

 \begin{figure}[tbhp]
\includegraphics[width=8cm,height=7cm]{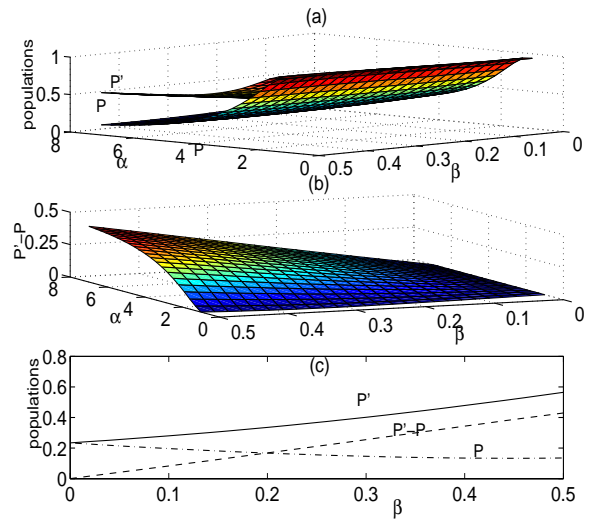}
\caption{(Color online) (a) The populations of atomic state $\vert
1\rangle$ versus variables $\alpha$ and $\beta$. $P$ $(P')$ is
the population of the state $|1\rangle$ at the end of the
evolution $U$ $(U^{'})$.
 (b) The difference $P'-P$ of the two populations
shown in (a) as a functions of the parameters $\alpha$ and
$\beta$. (c) The populations $P$, $P'$ and $P'-p$ versus the
parameter $\beta$ when $\alpha=7$.}
\end{figure}

\begin{figure}[tbhp]
\includegraphics[width=8cm,height=6cm]{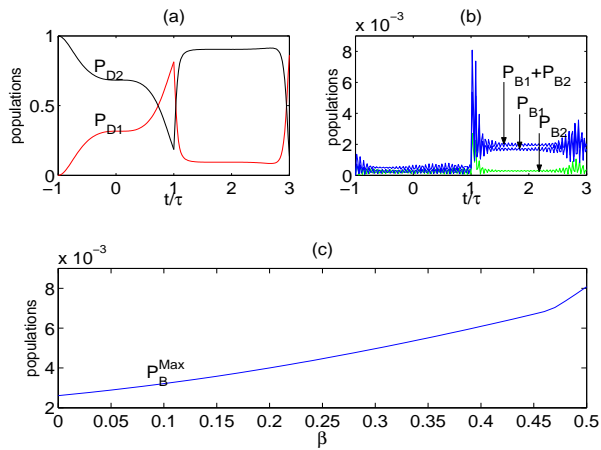}
\caption{(Color online) (a) The populations $P_{D1}$ ($P_{D2})$
for the dark state $\vert D_{1}\rangle$ ($\vert D_{2}\rangle$)
during a complete composite evolution $U$. (b) The populations
$P_{B1}$ $(P_{B2})$ for the bright state $\vert B_{1}\rangle$
$(\vert B_{2}\rangle)$. (c) The maxima $P_{B}^{Max}$ of the sum
$P_{B1}$ and $P_{B2}$ during the whole evolution time versus
parameters $\beta$. The parameters $\Omega_0=100$ and $\tau=2.$}
\end{figure}

We now turn to demonstrate that the adiabatic condition crucially
required in the proposed scheme is well satisfied.  Besides the
two degenerate dark states described in Eq.(\ref{Dark}), there
exist two other bright states with eigenvalues  $\pm \Omega$ in
the system described by the Hamiltonian (\ref{H_I}), and the
corresponding eigenstates are given by ,
\begin{eqnarray}
\vert B_{1}\rangle =&& \frac{1}{\sqrt{2}}[\sin\theta \cos \phi
e^{i\varphi}\vert 1 \rangle + \sin\theta \sin \phi e^{i\varphi}\vert 2 \rangle \notag\\
&&+ \cos\theta \vert 3\rangle +e^{i\varphi}\vert 0\rangle],\notag \\
\vert B_{2}\rangle =&& \frac{1}{\sqrt{2}}[\sin\theta \cos \phi e^{i\varphi}\vert 1 \rangle +
\sin\theta \sin \phi e^{i\varphi}\vert 2 \rangle\notag\\
&&+ \cos\theta \vert 3\rangle - e^{i\varphi}\vert 0\rangle].
\label{Bright}
\end{eqnarray}
The adiabatic approximation is well satisfied if the overflow from
the initial dark states to the two bright states (\ref{Bright}) can
be negligible during the whole evolution. In view of this point, we
numerically calculate the populations of all dark and bright states.
The results for the parameters $\alpha=7$ and $\beta=0.5$ are
plotted in Fig. 3(a) and (b). It is shown that the populations for
both bright states can be negligible. In particular, we calculate
the maxima ($P_{B}^{Max}$) of the sum $P_{B1}$ and $P_{B2}$ during
the whole evolution time versus the parameter $\beta$ with the other
parameters being the same as those in Fig.2(c). We note that the sum
of the populations for the two bright states are always below
$1.0\%$ during the whole evolution. Therefore it clearly shown that
the adiabatic approximation used in the calculations of data in
Fig.2 (c) is reasonable. Usually the adiabatic approximation is
well-satisfied when $\Omega_0\tau \gg 1$ \cite{Unanyan1999,
Bergmann}.

In summary, we have proposed an experimentally feasible scheme to
detect a true non-Abelian geometric phase effect in trapped ions
coupled with laser beams.
By designing two specific composite cyclic evolutions formed by
$U_{2}U_{1}$ and $U_{1}U_{2}$, we show in detail how the effect
induced by the noncommutativity of non-Abelian gauge structures can
be observed through detecting the population of the internal states
of trapped ions. The high-precision detection of the atomic states
recently developed in the field of quantum information opens the
great possibility to directly ``see'' the noncommutative feature of
the gauge potentials. It is quite promising for experimentalists to
implement the interest idea presented in this work.

This work was supported by the RGC of Hong Kong under Grant No.
HKU7045/05P and HKU7049/07P, the URC fund of HKU, the NSFC (Nos.
10429401, 10674049, and 60578055), NCET and the State Key Program
for Basic Research of China (Nos. 2006CB921800 and 2007CB925204).



\begin{thebibliography}{99}
\bibitem{Note} The scheme proposed here are equally well for
cold atoms since the effective Hamiltonian for an atom interacting
with three laser pulses may be exactly the same with the
Hamiltonian (\ref{H_I}).
\end{thebibliography}
\end{document}